\documentclass[traditabstract]{aa} 
\usepackage{graphicx}
\usepackage{txfonts}
\usepackage{color}

\newcommand\kms{{\rm\,km\,s^{-1}}}
\newcommand\msun{M_\odot}

\begin{document}
\titlerunning{Discovery of a parsec-scale bipolar nebula around MWC\,349A}
\authorrunning{Gvaramadze \& Menten}

   \title{Discovery of a parsec-scale bipolar nebula around MWC\,349A}

   \author{V.~V.~Gvaramadze
          \inst{1,2,3}
          \and
          K.~M.~Menten\inst{4}
          }

   \institute{Argelander-Institut f\"{u}r Astronomie, Universit\"{a}t Bonn, Auf dem H\"{u}gel 71, 53121 Bonn,
   Germany \\ \email{vgvaram@mx.iki.rssi.ru} \and Sternberg Astronomical Institute, Lomonosov Moscow State
     University, Universitetskij Pr. 13, Moscow 119992, Russia \and Isaac Newton Institute of Chile, Moscow Branch,
              Universitetskij Pr. 13, Moscow 119992, Russia \and
             Max-Planck-Institut f\"ur Radioastronomie, Auf dem H\"ugel 69, 53121 Bonn, Germany\\
             \email{kmenten@mpifr-bonn.mpg.de}
                      }

   \date{Received 18 January 2012; accepted 13 March 2012}

  \abstract
   {We report the discovery of a bipolar nebula around the
   peculiar emission-line star MWC\,349A using archival {\it Spitzer Space Telescope}
   24\,$\mu$m data. The nebula extends over several arcminutes (up to 5 pc)
   and has the same orientation and geometry as the well-known
   subarcsecond-scale ($\sim 400$ times smaller) bipolar radio nebula
   associated with this star. We discuss the physical relationship between MWC\,349A and the nearby B0\,III
   star MWC\,349B and propose that both stars were members of a hierarchical triple
   system, which was ejected from the core of the Cyg\,OB2 association several Myr ago
   and recently was dissolved into a binary system (now MWC\,349A) and a single unbound
   star (MWC\,349B). Our proposal implies that MWC\,349A is an evolved
   massive star (likely a luminous blue variable) in a binary system with a low-mass
   star. A possible origin of the bipolar nebula around MWC\,349A is discussed.}

   \maketitle

\section{Introduction}

MWC\,349A is a curious star located on the sky in the direction of
the Cyg\,OB2 association. It was identified as a peculiar
emission-line star in 1932 by Merrill et al. (1932), who also
noted the presence of a second star, MWC\,349B, $\approx 2$ arcsec
west. Since that time MWC\,349A attracted wide attention and
although many interesting details have been revealed about it, the
nature of this star remains puzzling.

MWC\,349A is one of the brightest radio stars on the sky (Braes et
al. 1972). Its radio emission was resolved with the Very Large
Array (VLA) in a sub-arcsecond square nebula by Cohen et al.
(1985), who suggested that the shape of the nebula might be due to
a biconical outflow of ionized gas pinched at the waist by a disk
seen edge-on and oriented due east-west. This suggestion was
confirmed by White \& Becker (1985), whose higher resolution and
higher frequency data clearly revealed a hourglass-shaped bipolar
nebula with a symmetry axis almost in the north-south direction
(see also Tafoya et al. 2004). MWC\,349A also known as a source of
hydrogen (sub)millimeter recombination lines that show maser
(Mart\'in-Pintado et al. 1989) and laser action (Strelnitski et
al. 1996). With increasing frequency, these lines attain a more
and more pronounced two-peaked profile that was interpreted as
originating from a disk or ring-shaped structure in Keplerian
rotation around the central star of mass of $\sim 30 \, \msun$
(Thum et al. 1992, Ponomarev et al. 1994). (Sub)millimeter
interferometry showed the emission to be consistent with a disk
(Planesas et al. 1992), but revealed a velocity structure not
consistent with Keplerian, which prohibits a mass estimate for the
central object (Weintroub et al. 2008).

Cohen et al. (1985) found a signature of interaction between the
nebula and MWC\,349B and suggested that this star forms a binary
system with MWC\,349A (see also Tafoya et al. 2004). Cohen et al.
(1985) were also able to separate the spectra of the two stars and
classified MWC\,349B as a B0\,III star. The spectrum of MWC\,349A
is dominated by a multitude of very strong emission lines, which
completely hide the photospheric lines (Cohen et al. 1985;
Andrillat et al. 1996) and make a spectral classification of the
star impossible. Using the spectral type of MWC\,349B, Cohen et
al. (1985) estimated a distance to the putative binary system of
1.2 kpc, which is much smaller than the distance to Cyg\,OB2
accepted at this time, i.e., 2 kpc. The physical relationship
between MWC\,349A and MWC\,349B was questioned by Meyer et al.
(2002). These authors measured polarization towards the two stars
and found that MWC\,349B is more highly polarized than MWC\,349A.
On the other hand, they found that several members of Cyg\,OB2
near to MWC\,349A have polarization characteristics similar to
those of the interstellar polarization towards this star. Based on
these findings, Meyer et al. (2002) suggested that MWC\,349A and
MWC\,349B are members of the Cyg\,OB2 association and that they
are projected by chance near the same line of sight, with
MWC\,349B located ``behind the dusty region around MWC\,349A".
Meyer et al. (2002), however, did not explain how to reconcile the
distances to Cyg\,OB2 and MWC\,349B. The problem of the physical
relationship between MWC\,349A and MWC\,349B is very important for
understanding the nature of the former star. If both stars form a
binary system, then MWC\,349A should be a massive evolved star,
while the distance to the system should be equal to the
spectroscopic distance to MWC\,349B. In Sect.\,\ref{sec:dis} we
argue that both stars could indeed reside in the Cyg\,OB
association and at the same time they could be physically related
to each other (at least in the recent past).

The bipolar radio nebula associated with MWC\,349A is produced by
a constant velocity outflow of ionized gas (Olnon 1975; Hartmann
et al. 1980). Widths of the [N\,{\sc ii}] $\lambda 6584$ emission
line (Hartmann et al. 1980) and the hydrogen radio recombination
lines (Altenhoff et al. 1981) suggest that the velocity of the
outflow is $\sim 50 \, \kms$. Hartmann et al. (1980) argued that
the low-velocity nature of the wind could be understood if the
star has a low effective surface gravity (i.e. the star is
inflated) and interpreted MWC\,349A as a supergiant star of
P\,Cygni type. Radio continuum observations indicate an outflow
mass-loss rate of $\sim 10^{-5} \, \msun {\rm yr}^{-1}$ (Hartmann
et al. 1980; Altenhoff et al. 1981; Dreher \& Welch 1983). With
this mass-loss rate and the low wind velocity the line emission in
the optical and the near-infrared must become optically thick if
the outflow originates at a radius of $\la 1$ AU (Altenhoff et al.
1981). Since the emission lines appear to be optically thin
(Thompson et al. 1977), it was suggested that the outflow is fed
by a circumstellar disk, so that the velocity of the wind is
rather set by the escape velocity from the disk than by that from
the stellar surface (Altenhoff et al. 1981). The presence of a
(nearly edge-on) circumstellar disk also allows to explain the
coexistence of low- and high-excitation emission lines in the
spectrum of MWC\,349A (Hamann \& Simon 1986), the double-peaked
profiles of optical, infrared (Hamann \& Simon 1986, 1988) and
masing (Planesas et al. 1992; Thum et al. 1992; Gordon 1992)
emission lines, and the high level of polarization in this star
(Elvius 1974; Yudin 1996; Meyer et al. 2002). The existence of the
disk was confirmed  with infrared speckle interferometry (Leinert
1986; Danchi et al. 2001; Hofmann et al. 2002), which revealed a
flattened structure (with a major axis of $\approx 60$ mas or
$\approx 85$ AU at our adopted distance of 1.4 kpc; see
Sect.\,\ref{sec:bir}) lying in the east-west plane, i.e. oriented
perpendicular to the symmetry axis of the bipolar radio nebula. It
is widely accepted that the disk plays a crucial role in formation
and shaping the radio nebula, whose extent along the polar axis
(as seen in the VLA 2 cm image; see Sect.\,\ref{sec:rad}) is
$\approx 1000$ AU or $\approx 0.005$ pc.

In this paper, we report the discovery of a parsec-scale
mid-infrared bipolar nebula around MWC\,349A, whose orientation
and geometry are similar to those of the sub-arcsecond radio
nebula. The close similarity of the morphologies of the two
nebulae suggests that they are shaped by the same agent -- the
disk around MWC\,349A. The newly-discovered nebula is presented in
Sect.\,\ref{sec:neb}. In Sect.\,\ref{sec:dis} we discuss the
location and the evolutionary status of MWC\,349A, the possible
origin of the bipolar outflow associated with this star, and the
similarity between this outflow and bipolar nebulae produced by
other massive evolved stars. We summarize in Sect.\,\ref{sec:sum}.

\section{Mid-infrared, radio and H$\alpha$ nebulae around MWC\,349A}
\label{sec:neb}

\subsection{{\it Spitzer} data}
\label{sec:inf}

The large-scale infrared nebula around MWC\,349 was detected in
archival data originating from observations of the {\it Spitzer
Space Telescope}, namely within the framework of the Cygnus-X
Spitzer Legacy Survey (Hora et al.
2008)\footnote{http://www.cfa.harvard.edu/cygnusX}. This survey
covers 24 square degrees in Cygnus\,X, one of the most massive
star-forming complexes in the Milky Way (e.g., Piddington \&
Minnett 1952; Reipurth \& Schneider 2008), and provides images at
24 and 70\,$\mu$m obtained with the Multiband Imaging Photometer
for Spitzer (MIPS; Rieke et al. 2004) and at 3.6, 4.5, 5.8, and
8.0\,$\mu$m obtained with the Infrared Array Camera (IRAC; Fazio
et al. 2004). The resolution of the MIPS 24 and 70\,$\mu$m images
is $\approx 6$ and 18 arcsec, respectively, while that of the IRAC
images is $\approx 1$ arcsec. Inspection of the data from this
survey have already led to the discovery of a new Wolf-Rayet star
(through the detection of a circular 24\,$\mu$m shell and
follow-up spectroscopy of its central star; Gvaramadze et al.
2009) and a second (concentric) shell around the already known
infrared shell (Trams et al. 1998; Egan et al. 2002) surrounding
the candidate luminous blue variable (cLBV) star GAL\,079.29+00.46
(Gvaramadze et al. 2010a; Jim\'{e}nez-Esteban et al. 2010; Kraemer
et al. 2010). The MIPS and IRAC post-basic calibrated data on
MWC\,349A were retrieved from the NASA/IPAC Infrared Science
Archive\footnote{http://irsa.ipac.caltech.edu/}. Like many other
circumstellar nebulae discovered with {\it Spitzer} (e.g.
Gvaramadze et al. 2010a; Wachter et al. 2010; Mizuno et al. 2010),
the nebula associated with MWC\,349A is best visible at
$24\,\mu$m. It can also be seen at 70\,$\mu$m, but in none of the
IRAC images.

\begin{figure*}
 \resizebox{18cm}{!}{\includegraphics{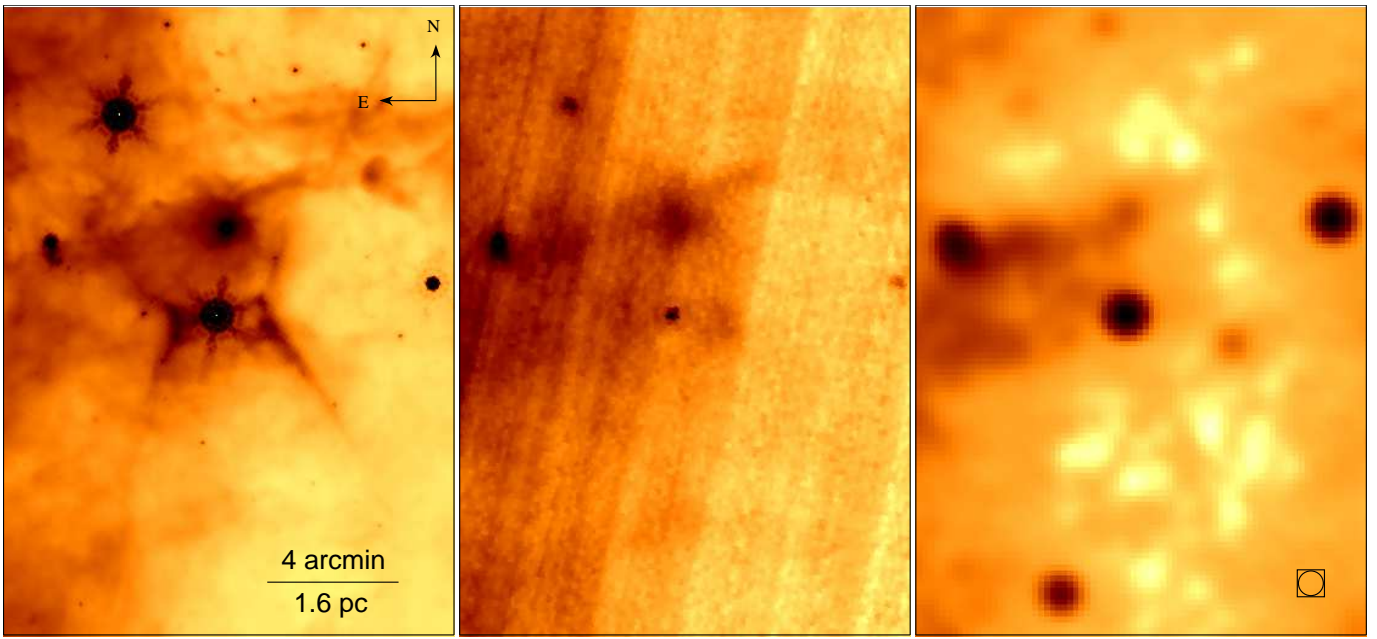}}
 \caption{{\it Left}: MIPS $24\,\mu$m image of the arcminute-scale bipolar nebula
 and its central star MWC\,349A (a bright source in the waist of the nebula; note that a white
 dot at the centre of this source is due to saturation effect). {\it Middle} and {\it Right}:
 MIPS $70\,\mu$m and VLA 17 cm images of the same field. The $47''$ FWHM restoring beam is
 represented in the bottom right corner of the radio image.
 }
  \label{fig:neb}
\end{figure*}

The left and the middle panels of Fig.\,\ref{fig:neb} present the
MIPS 24 and 70\,$\mu$m images of the nebula and its central star
MWC\,349A (a white dot at the centre of the star in the MIPS\,24
$\mu$m image is due to saturation effect). At 24\,$\mu$m the
nebula has a pronounced bipolar (hourglass-like) structure
stretched almost in the north-south direction. The waist of the
nebula is surrounded by a bright belt with a diameter of $\approx
2.7$ arcmin. The nebula is somewhat asymmetric with respect to the
polar axis. From the west side the lobes are outlined by almost
straight filaments, with the northern and the southern ones
extended, respectively, out to $\approx 5.5$ and 6 arcmin from the
waist. From the east side, the lobes are bounded by somewhat
curved filaments. The northern filament extends out to $\approx 3$
arcmin from  the waist and then merges with the bright background,
while the southern one can be traced up to $\approx 7$ arcmin. The
curvature of the latter filament suggests that the southern lobe
might be a closed structure.

At $\approx 2$ arcmin from the waist the eastern filament of the
southern lobe splits into two parts and becomes concave towards
the interior of the lobe, which might indicate an interaction
between the laterally expanding nebula and the ambient medium. At
our adopted distance of 1.4 kpc (see Sect.\,\ref{sec:bir}), 1
arcmin corresponds to $\approx 0.4$ pc, so that the diameter of
the waist is $\approx 1.1$ pc, while the lobes stretch out to
$\approx 2.4-2.8$ pc from the equatorial belt. Interestingly,
MWC\,349A is offset by $\approx 20$ arcsec (or $\approx 0.14$ pc)
from the geometric centre of the waist, being closer to its
eastern edge. This displacement could be understood if the nebula
impinges on a more dense ambient medium in the eastern direction.
The presence of the dense material on the eastern side of the
nebula might be inferred from the $70\,\mu$m and the 17 cm radio
wavelength (see Sect.\,\ref{sec:rad}) images of the field around
the nebula (see the middle and the right panels of
Fig.\,\ref{fig:neb}, respectively). The $70\,\mu$m image also
shows obvious counterparts to the equatorial belt and the eastern
boundary of the southern lobe.

\subsection{VLA data}
\label{sec:rad}

To produce a long wavelength (L-band, $\approx 17$ cm) radio image
of MWC\,349A, we used two archival datasets obtained with the NRAO
Very Large Array (VLA)\footnote{The VLA is operated by the
National Radio Astronomy Observatory (NRAO). The NRAO is a
facility of the National Science Foundation operated under
cooperative agreement by Associated Universities, Inc.}. The
datasets were acquired and calibrated by R. Perley in the course
of a long range monitoring campaign. The data was taken on 2008
August 31 in the most compact (D) configuration at 1665.9 and
1875.1 MHz, each with one intermediate frequency band with
bandwidth 12.5 MHz. The data were processed further with programs
of the Astronomical Image Processing System (AIPS) in the usual
manner. First, the two datesets were combined (using DBCON) and
then imaged with IMAGR. One cycle of self calibration (with CALIB)
improved the image quality. The image, part of which is presented
in Fig.\,\ref{fig:neb}, was restored with a circular Gaussian beam
with FWHM $47''$. There are no clear indications of the presence
of radio emission associated with the mid-infrared nebula.

\begin{figure}
 \resizebox{8cm}{!}{\includegraphics{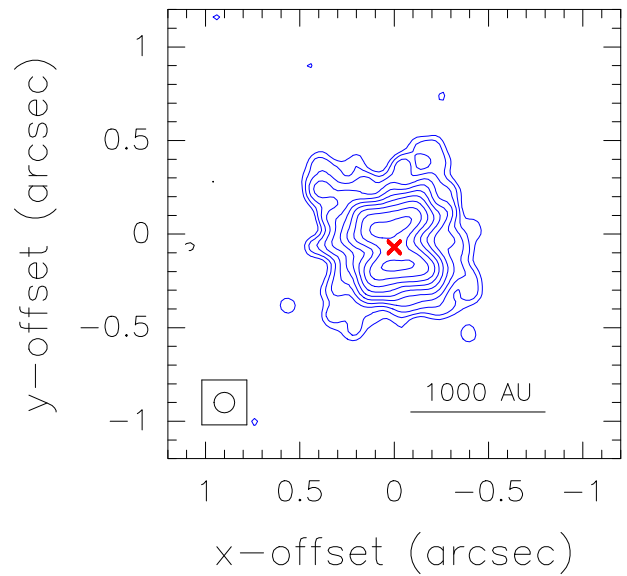}}
 \caption{VLA 2 cm image of the subarcsecond bipolar nebula around
 MWC\,349A (indicated by a cross). The angular size of the field
 shown in this figure is equal to that of a white spot in the centre of the 24\,$\mu$m image
 of MWC\,349A (see the left panel of Fig.\,\ref{fig:neb}). The $0\rlap{.}''11$ FWHM restoring
 beam is represented in the bottom left corner.}
  \label{fig:VLA}
\end{figure}

We also produced a radio image of MWC\,349A with higher resolution
taken in the most extended (A) configuration at a shorter
wavelength (U-band, 2 cm) from archival data taken on 1983 October
30 at a frequency of 14.94 GHz. The observations are reported and
discussed by White \& Becker (1985) and we refer to this
publication for details on the data and their processing.
Additional use of self-calibration resulted in an improved image.
This image (presented in Fig.\,\ref{fig:VLA}) shows the well-known
subarcsecond-scale radio nebula associated with MWC\,349A.
Comparison of Fig.\,\ref{fig:VLA} with the MIPS 24\,$\mu$m image
of the arcminute-scale nebula shows that both nebulae have the
same orientation and bipolar morphology. Note that by coincidence
the angular size of the field shown in Fig.\,\ref{fig:VLA} turns
out to be equal to that of the saturated pixel (the white dot) in
the very centre of the 24\,$\mu$m image of MWC\,349A.

\subsection{H$\alpha$ data}
\label{sec:Ha}

The parsec-scale nebula around MWC\,349A was also detected in the
H$\alpha$ narrow-band CCD imaging survey of B[e] stars by Marston
\& McCollum (2008). These authors noted that ``an apparent thin
shell of approximately $2\farcm5$ diameter exists around MWC\,349A
to the north". Comparison of their Fig.\,4a with the 24\,$\mu$m
image in Fig.\,\ref{fig:neb} shows that a thin H$\alpha$ filament
west of MWC\,349A coincides with the western edge of the
equatorial belt, while an arc-like diffuse H$\alpha$ emission to
the east shows a good correspondence with the eastern edge of the
waist.

\section{Discussion}
\label{sec:dis}

Above we have reported the discovery of a parsec-scale
mid-infrared nebula around the enigmatic emission-line star
MWC\,349A, whose hourglass-like structure and orientation are
similar to those of the well-known subarcsecond radio nebula
associated with this star. The morphology of the nebula is very
similar to that of some planetary nebulae (e.g. the Engraved
Hourglass Nebula; Sahai et al. 1999), nebulae around symbiotic
systems (e.g. the Southern Crab Nebula; Corradi \& Schwarz 1993),
and nebulae associated with young stellar objects (e.g. the
disk/outflow radio source I in the Kleinmann-Low Nebula in Orion;
Plambeck et al. 2009; Matthews et al. 2010) and some evolved
massive stars [e.g. the cLBV Sher\,25 (Brandner et al 1997) and
the famous LBV $\eta$\,Carinae]. Correspondingly, three main
interpretations of the subarcsecond bipolar nebula associated with
MWC\,349A were proposed: (i) planetary nebula (Ciatti et al.
1974), (ii) photo-evaporation induced outflow from an accretion
disk around a massive pre-main sequence star (e.g. Thompson et al.
1977), and (iii) outflow from an (excretion) disk around an
evolved massive star (Hartmann et al. 1980).

Hofmann et al. (2002) thoroughly analyzed these three
possibilities and came to the conclusion that MWC\,349A is most
likely an evolved massive star (a B[e] supergiant) residing in the
Cyg\,OB2 association. Similar proposals on the evolutionary status
of MWC\,349A were also put forward by Swings \& Struve (1942),
Baldwin et al. (1973), Hartmann et al. (1980), Herzog et al.
(1980), Andrillat et al. (1996), and Lamers et al. (1998). The
membership of the Cyg\,OB2 association seems, however, to be at
variance with the possibility that MWC\,349A forms a physical
system with MWC\,349B (Cohen et al. 1985; Tafoya et al. 2004),
since the spectroscopic distance to the latter star of $\sim 1.2$
kpc is significantly smaller than the today's widely accepted
distance to Cyg\,OB2 of 1.7 kpc (Kn\"{o}dlseder 2000; Reipurth \&
Schneider 2008). Before discussing the possible origin of the
bipolar nebula associated with MWC\,349A, we have to understand
whether the physical relationship between the two stars could be
consistent with their membership of the Cyg\,OB2 association.

\subsection{Birth cluster and the evolutionary status of MWC\,349A}
\label{sec:bir}

Let us assume that MWC\,349A is a massive evolved star as argued
by Hofmann et al. (2002). In this case, the question arises: where
is the birth cluster of this star? Indeed, there is growing
acceptance that most (or probably all) massive stars are formed in
the clustered way (Lada \& Lada 2003) and subsequently found
themselves in the field either because of dynamical few-body
encounters in the parent clusters, binary-supernova explosions
and/or due to rapid cluster dissolution (e.g., Kroupa \& Boily
2002; de Wit et al. 2005; Schilbach \& R\"oser 2008; Gvaramadze \&
Bomans 2008b; Pflamm-Altenburg \& Kroupa 2010; Gvaramadze et al.
2010b; Weidner et al. 2011)\footnote{See Parker \& Goodwin (2007),
Selier, Heydari-Malayeri \& Gouliermis (2011), and references
therein for a different point of view.}. One can therefore expect
that the parent cluster of MWC\,349A should be nearby, unless this
star is a high-velocity runaway. Since MWC\,349A is located in the
direction of one of the most compact and massive associations in
the Milky Way -- the Cyg\,OB2 association (Kn\"{o}dlseder 2000),
it is natural to assume that it is a member of this stellar
system.

If MWC\,349A is indeed a member of Cyg\,OB2 then the small angular
separation between MWC\,349A and MWC\,349B might simply be due to
a chance projection. The line of sight towards MWC\,349A is nearly
tangential to the local (Orion) spiral arm (whose extent in this
direction is $\sim 4-6$ kpc; e.g. Russeil 2003), which makes the
chance projection quite possible. The SIMBAD
database\footnote{http://simbad.u-strasbg.fr/simbad/} lists three
O stars within 10 arcmin from MWC\,349A, one of which, the O8\,III
(Negueruela et al. 2008) star 2MASS J20323843+4040445, is
separated from MWC\,349A by only $\approx 1.8$ arcmin and is
projected on the western edge of the 24\,$\mu$m nebula (not far
from the equatorial belt). We note, however, that recent very
accurate parallax measurements of five massive star-forming
regions towards Cygnus\,X showed that four of them are located at
a distance of $1.4\pm 0.1$ kpc (Rygl et al. 2012). At the same
time, there are strong indications that most molecular clouds in
Cygnus\,X form a coherent complex located at the same distance as
Cyg\,OB2 (Schneider et al. 2006), which implies that Cyg\,OB2 is
at the distance of $1.4\pm0.1$ kpc as well (cf. Hanson 2003). From
this it follows that the physical relationship between MWC\,349A
and MWC\,349B and their membership of Cyg\,OB2 do not contradict
each other, which in turn implies that MWC\,349A should be as old
as MWC\,349B (i.e. $\sim 5$ Myr). The obvious consequence of the
latter implication is that MWC\,349A cannot be a young stellar
object, nor a post-AGB star.

To check whether the spectral classification of MWC\,349B is
consistent with the possibility that this star is a member of
Cyg\,OB2, we use its visual magnitude and extinction of,
respectively, $V=14.3$ mag and $A_V =8.6-8.9$ mag (Cohen et al.
1985) to derive the absolute visual magnitude $M_V = -(5.0\div
5.3)$ mag, which agrees well with that expected for a B0\,III
star.

Similarly, assuming the visual magnitude of MWC\,349A of 14 mag
(Cohen et al. 1985) and adopting $A_V =10.0-10.6$
mag\footnote{Note that the interstellar extinction towards the
Cyg\,OB2 association is very patchy and ranges from $\sim 5$ to 20
mag (Kn\"{o}dlseder 2000).} (Cohen et al. 1985; Kelly et al.
1994), one has $M_V = -(6.7\div7.3)$ mag. The absence of He\,{\sc
ii} lines and the presence of strong He\,{\sc i} emission lines in
the spectrum of MWC\,349A imply that the effective temperature of
this star should be in a range from $\sim 20\,000$ to 28\,000 K,
which corresponds to a bolometric correction of $-2.5\pm 0.4$ mag
(Hofmann et al. 2002). Thus, one derives a the stellar luminosity
of $\log (L/L_{\odot})=(5.6\div5.9)\pm 0.2$ (cf. Hartmann et al.
1980; Hofmann et al. 2002). We caution that this luminosity
estimate should be considered approximate because of the large
($\approx 1-2$ mag) photometric variability of MWC\,349A: in the
$B$-band its magnitude varies from $\approx 14$ to 16 mag
(Gottlieb \& Liller 1978), while in the $R$-band from $\approx
9.5$ to 10.5 mag (Jorgenson et al. 2000). On the other hand, the
good correlation between the brightness in the $V$ and $R$ bands
in MWC\,349A (see Fig.\,1 in Yudin 1996) implies that $V=14$ mag
corresponds to the minimum brightness of this star, so that its
actual luminosity could be even higher.

The high luminosity of MWC\,349A, along with its significant
spectroscopic (e.g. Andrillat et al. 1996) and photometric
variability, and the presence of the parsec-scale (bipolar)
circumstellar nebula are typical of evolved massive stars
belonging to the class of LBVs (e.g. Humphreys \& Davidson 1994;
Bohannan 1997). We therefore concur with Hofmann et al. (2002)
that MWC\,349A is an evolved massive star, likely an LBV (cf.
Hartmann et al. 1980). This inference is supported by the possible
existence of a high-velocity ($\approx 260 \, \kms$) component in
the bipolar outflow (Tanaka et al. 1985), which might correspond
to the LBV wind concentrated along the polar axis (we will return
to this point in Sect.\,\ref{sec:ori}). Assuming that MWC\,349A is
an LBV of age of 5 Myr and using the above estimate of its
luminosity, one can infer the zero-age main-sequence mass of this
star of $\sim 40 \, \msun$.

\subsection{MWC\,349A as a runaway}
\label{sec:run}

MWC\,349A is separated from the geometric centre of the Cyg\,OB2
association by $\approx 24$ pc in projection. Cyg\,OB2 contains
$\sim 100$ O stars or stars with O-type progenitors and its half
light radius is $\sim 6$ pc (Kn\"odlseder 2000). The age of the
association is $\sim 5$ Myr (see Gvaramadze \& Bomans 2008a and
references therein). Assuming that the association expands with a
velocity equal to its velocity dispersion ($\approx ­2.4 \, \kms$;
Kiminki et al. 2007), one finds that the majority of massive stars
in Cyg\,OB2 were originally concentrated in a region of radius of
$<1$ pc. It is therefore plausible that several Myr ago the
stellar number density in the core of Cyg\,OB2 was high enough to
ensure that close dynamical encounters between its members were
frequent, which is the necessary condition for effective
production of runaways. Currently only one runaway associated with
Cyg\,OB2 is known -- the O4\,If star BD+43$\degr$\,3654 (Comer\'on
\& Pasquali 2007; Gvaramadze \& Bomans 2008a). Let us check
whether MWC\,349A might be another example of a massive star
running away from Cyg\,OB2.

To check this possibility, we use the proper motion measurements
for MWC\,349A given in Rodr\'{i}guez et al. (2007). To convert the
observed proper motion ($\mu _\alpha \cos \delta =-3.1\pm 0.5$ mas
${\rm yr}^{-1}$ and $\mu _\delta =-5.3\pm 0.5$ mas ${\rm
yr}^{-1}$) into the transverse peculiar velocity, we use the
Galactic constants $R_0 = 8.0$ kpc and $\Theta _0 =240 \, {\rm km}
\, {\rm s}^{-1}$ (Reid et al. 2009) and the solar peculiar motion
$(U_{\odot} , V_{\odot} , W_{\odot})=(11.1, 12.2, 7.3) \, {\rm km}
\, {\rm s}^{-1}$ (Sch\"{o}nrich et al. 2010). The derived velocity
components in Galactic coordinates are $v_l = -7.3\pm6.8 \, \kms$
and $v_b =3.0\pm3.3 \, \kms$ (for the error calculation, only the
errors of the proper motion measurements and the distance estimate
were considered). With the transverse peculiar velocity of $v_{\rm
tr} \approx 8
\, \kms$, MWC\,349A would need $\approx 3$ Myr to travel from the
centre of the association to its current position.

\begin{figure}
\begin{center}
\includegraphics[width=9cm,angle=0,clip=]{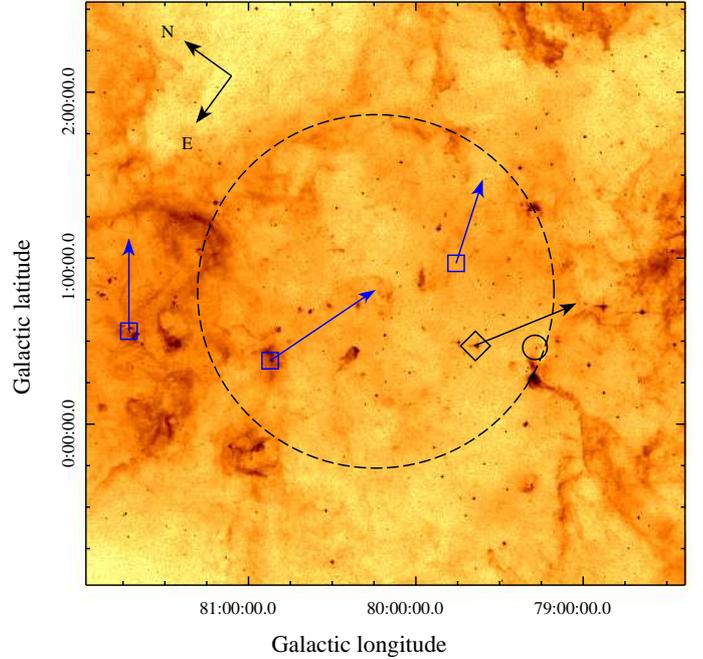}
\end{center}
\caption{{\it Midcourse Space Experiment} ({\it MSX}) 8.3~$\mu$m
image of the Cyg\,OB2 association and its environments. The
approximate boundary of the association is shown by a dashed
circle of a diameter of $\approx 2\degr$. The position of
MWC\,349A is marked by a diamond, while that of the nearby cLBV
GAL\,079.29+00.46 is indicated by a circle. The arrows show the
space motions of MWC\,349A and three star-forming regions
(indicated by squares). See text for details.} \label{fig:CygOB2}
\end{figure}

Figure\,\ref{fig:CygOB2} shows that the peculiar velocity of
MWC\,349A has somewhat ``wrong" orientation, i.e. it does not
point away from the centre of the Cyg\,OB2 association. To explain
this misalignment, one can assume that Cyg\,OB2 has a peculiar
(transverse) velocity of $\sim 5-10 \, \kms$ in the northwest
direction (cf. Hoogerwerf et al. 2001). The peculiar velocity of
just this magnitude and orientation is required to explain the
relative position of the runaway star BD+43$\degr$\,3654 and two
young pulsars on the sky (Gvaramadze \& Bomans 2008a). Peculiar
velocities of $\sim 10 \, \kms$ are typical of the OB associations
near the Sun (de Zeeuw et al. 1999). More importantly, three
star-forming regions adjacent to the Cyg\,OB2 association (DR20,
DR21 and IRAS 20290+4052) have peculiar (transverse) velocities of
$\approx 6-9 \, \kms$ directed northwest and west (Rygl et al.
2012; see also Fig.\,\ref{fig:CygOB2}).

To the peculiar transverse velocity of MWC\,349A one should add
the peculiar radial velocity, $v_{\rm rad}$, which can be derived
from the recession velocity of this star of $8-12 \, \kms$ (e.g.
Thum et al. 1992; Gordon et al. 2001; Gordon 2003). After
correction for differential Galactic rotation and solar peculiar
motion, one has $v_{\rm rad} \approx 18-22 \, \kms$ and the total
space velocity $v_{\rm tot} \approx 20-23 \, \kms$. Although this
velocity does not meet the usual criterion to define runaway stars
(i.e. $v_{\rm tot} > 30 \, \kms$; Blaauw 1961), we note that the
ejection of stars with low velocities is common (e.g. Gies 1987;
Kroupa 1998; Gvaramadze \& Bomans 2008b), so that any star unbound
from the parent cluster should be considered as a "runaway",
independently of its peculiar velocity.

On the other hand, the derived total velocity of MWC\,349A is too
large to be consistent with the observed linear separation between
MWC\,349A and MWC\,349B ($\approx 3300$ AU) and the possibility
that these stars form a bound system. This inference follows from
the results of numerical scattering experiments by Kroupa (1998),
who showed that a wide binary system can survive the ejection
process only if its space velocity is less than the orbital
velocity. Simple estimates show that for any reasonable masses of
MWC\,349A and MWC\,349B the orbital velocity of the binary is an
order of magnitude smaller than its space velocity. A single
exception (i.e. a high-velocity wide binary) that occurred in the
experiments by Kroupa (1998) was interpreted as the outcome of ``a
complex high-order interaction, that resulted in two stars being
ejected on essentially identical trajectories". Another
possibility to produce a wide runaway binary is through the
ejection of a compact, initially stable hierarchical triple
system, which in the course of evolution of its components becomes
unstable and dissolves into a binary star and a single unbound
star. We further discuss this possibility in Sect.\,\ref{sec:ori}
because it might be directly related to the formation of the
bipolar nebula around MWC\,349A.

To conclude, we note that MWC\,349A is located at only $\approx
20$ arcmin (or $\approx 8$ pc in projection) from the cLBV
GAL\,079.29+00.46 (see Fig.\,\ref{fig:CygOB2}). The distance
estimates for GAL\,079.29+00.46 (Voors et al. 2000) suggest that
this star could be a member of Cyg\,OB2 as well. GAL\,079.29+00.46
is surrounded by a curious circumstellar nebula consisting of two
concentric circular shells (see, e.g., Fig.\,2j in Gvaramadze et
al. 2010a)\footnote{For another example of a two-shell
circumstellar nebula produced by the cLBV MN112 see Gvaramadze et
al. (2010c).}. The diameter of the inner (bright) shell is
$\approx 1.4$ pc (at the assumed distance to GAL\,079.29+00.46 of
1.4 kpc), while that of the outer shell is $\approx 2.8$ pc. We
speculate that the nebula around GAL\,079.29+00.46 might actually
have a bipolar geometry (similar to that around MWC\,349A) with
the polar axis parallel to our line of sight. In this case, the
inner shell would correspond to the circular waist of the nebula
around MWC\,349A (whose diameter is $\approx 1.1$ pc), while the
outer one to the bipolar lobes (whose diameter is $\approx 2.4$
pc). If subsequent studies of MWC\,349A will prove its LBV nature,
then along with GAL\,079.29+00.46 it would represent one more
example of massive stellar twinning (Walborn \& Fitzpatrick 2000).
Two other well-known examples of such twinning are the pairs of
(c)LBVs: HD\,168607 and HD\,168625 (separated by only $\approx
1.6$ arcmin) and AG\,Car and Hen\,3-519 (separated by $\approx
16.2$ arcmin).

\subsection{Possible origin of the bipolar nebula around MWC\,349A}
\label{sec:ori}

\begin{figure}
 \resizebox{9cm}{!}{\includegraphics{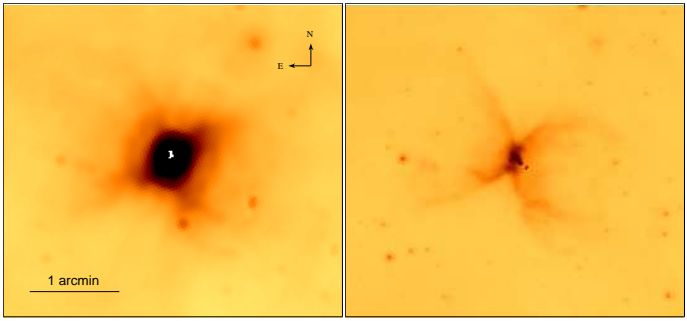}}
 \caption{MIPS 24\,$\mu$m (left) and IRAC $8\,\mu$m images of the bipolar nebula around
 the candidate LBV MN13 (Gvaramadze et al. 2010a). The orientation
 and the scale of the images are the same.}
  \label{fig:MN13}
\end{figure}

\begin{figure}
 \resizebox{9cm}{!}{\includegraphics{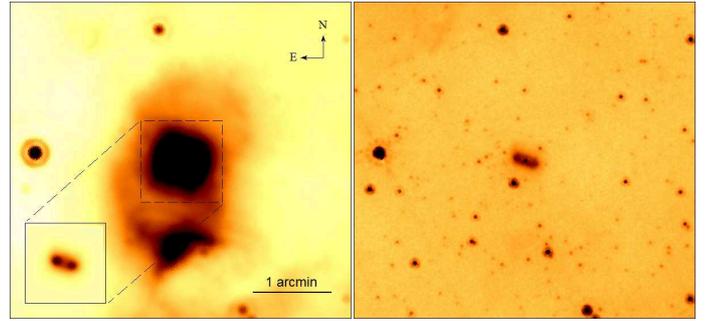}}
 \caption{MIPS 24\,$\mu$m (left) and IRAC $8\,\mu$m images of the
 bipolar nebula around the blue supergiant MN18 (Gvaramadze et al.
 2010a). The inset shows the waist of the nebula in a different
 intensity scale. The orientation and the scale of the images are
 the same.}
  \label{fig:MN18}
\end{figure}

In Sect.\,\ref{sec:bir} we suggested that MWC\,349A is likely an
LBV star. Observations show that almost all LBVs are surrounded by
compact nebulae with linear extent ranging from $\sim 0.1$ to
several pc (e.g. Weis 2001; Clark et al. 2005). These nebulae
display a wide diversity of shapes, from circular to bipolar and
triple-ring forms (e.g. Nota et al. 1995; Smith 2007; Gvaramadze
et al. 2010a). As a rule, the youngest LBV nebulae have bipolar
morphology\footnote{Note that such nebulae may become more
spherical with time because of the lateral expansion of their
polar lobes (cf. Sect.\,\ref{sec:mach}).} (see Table\,1 in Weis
2011). Three well-known examples of young ($\la 10^4$ yr) bipolar
nebulae associated with Galactic (c)LBVs are the Homunculus nebula
of the LBV $\eta$\,Car (e.g. Morse et al. 1998), the two-lobe
nebula around the cLBV Sher\,25 (Brandner et al. 1997), and the
triple-ring system around the cLBV HD\,168625 (Smith 2007). Two
other striking examples of hourglass-like nebulae produced by
evolved massive stars were recently discovered with {\it Spitzer}
(Gvaramadze et al. 2010a). One of them (Fig.\,\ref{fig:MN13}) is
created by the cLBV MN13 (Gvaramadze et al. 2010a; Wachter et al.
2011) and the second one (Fig.\,\ref{fig:MN18}) by the blue
supergiant star MN18 (Gvaramadze et al. 2010a). The MIPS
24\,$\mu$m image of the nebula around MN13 shows that the
brightest part of this nebula is reminiscent of the triple-ring
system produced by the cLBV HD\,168625 (see Fig.\,1d in Smith
2007), while the IRAC 8\,$\mu$m image shows that the waist of the
nebula is surrounded by a belt similar to that of the 24\,$\mu$m
nebula around MWC\,349A (the possible origin of such belts is
discussed in Sect.\,\ref{sec:mach}). The MIPS and IRAC images of
MN18 show that the central star is surrounded by a toroidal
structure, which most likely is responsible for the pinching the
waist of the nebula.


The widely accepted explanation of the origin of two-lobe nebulae
around evolved massive stars is that the (spherically symmetric)
stellar wind is collimated by an aspherical (disk-like)
circumstellar environment\footnote{Another possibility is that the
stellar wind is intrinsically bipolar (Owocki \& Gayley 1997;
Maeder \& Desjacques 2001).} (e.g. Frank et al. 1995; Langer et
al. 1999). Such structures could form around single massive stars
if they reach the $\Omega$ limit (i.e. if the centrifugal force
and the radiative acceleration cancel out gravity at the stellar
equator), so that the stellar wind becomes strongly confined to
the equatorial plane (Langer 1997, 1998). Alternatively, the
origin of disk-like density enhancements could be caused by
various binary interaction processes, such as (i) common envelope
ejection (e.g. Morris 1981), (ii) mass loss through the second
Lagrange point $L_2$ (e.g. Livio, Salzman \& Shaviv 1979), and
(iii) focusing of the primary's wind towards the orbital plane by
the gravitational field of the companion star (e.g. Fabian \&
Hansen 1979; Mastrodemos \& Morris 1999). All of these processes
can produce outflowing (excretion) disks in the orbital plane.

In Sect.\,\ref{sec:run} we mentioned that the large linear
separation between MWC\,349A and MWC\,349B could be reconciled
with their physical relationship if both stars were originally
members of a hierarchical triple system. In support of this
possibility we note that Jorgenson et al. (2000) reported
detection of a 9-year periodicity in the red light variations of
MWC\,349A, which implies that the star itself might be a binary
system (Jorgenson et al. 2000; Hofmann et al. 2002). Below we show
how an initially stable runaway triple system could be dissolved
into a binary system and a single unbound star, and suggest that
the gravitational focusing of the slow wind emanating from
MWC\,349A during the red supergiant phase might be responsible for
the origin of disk-like circumstellar environment, which in turn
is responsible for the origin and shaping of the observed bipolar
nebula.

Imagine that the inner binary of the triple system was composed of
MWC\,349A (with the zero-age main-sequence mass, $m_1$, of $40 \,
\msun$) and a low-mass star (with the mass, $m_2$, of, say, $5 \,
\msun$), while MWC\,349B (with the mass $m_3 =20\, \msun$) was in
a wide orbit. For the sake of simplicity, we assume that initially
both orbits were circular. This triple system would be hard enough
to survive dynamical ejection from Cyg\,OB2 if the orbital
separation of the outer binary $a_{\rm out} \la 100$ AU, while the
system itself would be stable if the ratio of the outer to inner
orbital periods $X\equiv P_{\rm out} /P_{\rm in} \ga 4$ (Kiseleva
et al. 1994), i.e. if the orbital separation of the inner binary
$a_{\rm in} < 35$ AU. A star with initial mass of $40 \, \msun$
evolves through the red supergiant and yellow hypergiant phases
(e.g. Humphreys 1991; Oudmaijer et al. 2009) and at the beginning
of the LBV phase has a mass of $\approx 20 \, \msun$. Adopting
this mass as the current mass of MWC\,349A, one has the current
mass of the inner binary of $25 \, \msun$, so that the 9-year
orbital period of this binary corresponds to an orbital separation
of $\approx 13$ AU (i.e. the secondary star orbit within the disk;
cf. Jorgenson et al. 2000). The large orbital separation implies
that the inner binary did not experience the common envelope
phase, i.e. the mass decrease of the primary star (MWC\,349A)
caused by (spherically symmetric) stellar wind mass loss led only
to the increase of the orbital period and separation of the binary
by factors of $(45/25)^2$ and 45/25, respectively. Thus, the
initial orbital period and separation of the inner binary were
$\approx 3$ yr and $\approx 8$ AU, respectively.

Let us assume that initially the triple was stable (i.e. $X>4$).
In the process of ejection from Cyg\,OB2 the outer binary acquired
an eccentricity $e_{\rm out}$ (cf. Hoffer 1983; Hills 1975) and
the perturbed triple system remained stable provided that
\begin{equation} e_{\rm out} \la 1-{F(x)\over X^{2/3}} \, ,
\label{eqn:ecc}
\end{equation}
where
\begin{equation}
F(x)={2.4x^2 +1.1x +3.7 \over (1+x^3)^{1/3} (1+x)}
\label{eqn:F}
\end{equation}
and $x=[(m_1 +m_2 )/m_3]^{1/3}$ (Eggleton \& Kiseleva 1995; Livio
\& Pringle 1998). Assuming $X=10$ (i.e. $P_{\rm out} =30$ yr and
$a_{\rm out} \approx 40$ AU), one finds from Eqs.\,(\ref{eqn:ecc})
and (\ref{eqn:F}) that the system is stable if $e_{\rm out} \la
0.42$. We assume that $e_{\rm out}$ obeys this condition (say,
$e_{\rm out} = 0.30$), so that the triple remains intact until the
primary of the inner binary, MWC\,349A, evolved off the main
sequence and lost a half of its initial mass during the red
supergiant and yellow hypergiant phases. In response to the mass
loss the orbital periods of the outer and the inner binaries
increased, respectively, by factors of $(65/45)^2 =2.09$ and
$(45/25)^2 = 3.24$ (we assume that $m_2$ and $m_3$ are constant),
so that $X$ decreased by a factor of 0.64 and the condition for
stability of the triple becomes $e_{\rm out} \la 0.19$. The
spherically symmetric mass loss does not affect the eccentricity
of the (inner and outer) binaries (e.g. Eggleton 2005). This and
our assumption of $e_{\rm out} =0.30$ implies that reduction of
the mass of MWC\,349A made the system unstable. According to the
numerical experiments by Kiseleva et al. (1994), an unstable
triple dissolves within $\sim 100$ crossing times (which in our
case corresponds to $\sim 10^4$ yr) and leaves behind a binary
system and a single unbound star. Thus, we propose that this
binary system corresponds to MWC\,349A and its putative low-mass
companion, while MWC\,349B is the third component of the triple,
which became unbound in the recent past.

We speculate that the slow (red supergiant) wind from MWC\,349A
was focused by the gravitational field of the companion star to
produce a circumbinary disk (see, e.g., Mastrodemos \& Morris
1999), which allows the current fast (LBV) wind of MWC\,349A to
expand only in the polar directions and is responsible for the
B[e] morphology of the stellar spectrum. We speculate also that
the hourglass-shaped low-velocity outflow observed as the
subarcsecond radio nebula originates because of gas entrainment in
the boundary layer between the stellar wind and the disk, while
the wind itself is confined near the polar axis of the slow
outflow and is mostly unobservable. The possible existence of a
high-velocity ($\approx 260 \, \kms$) component in the bipolar
radio nebula (Tanaka et al. 1985) conforms with our scenario and
could be attributed to the stellar wind channelled along the polar
axis. The energy injected in the ambient medium by the stellar
wind leads to the origin of parsec-scale laterally expanding
lobes, which we observe as the mid-infrared hourglass nebula. For
a wind velocity of $260 \, \kms$ and a linear extent of the lobes
of $\approx 3$ pc, one derives a kinematic age of the nebula of
$\approx 10^4$ yr. Note that this time-scale is comparable to the
dissolution time of the triple system (see above), so that it is
likely that MWC\,349B became unbound from MWC\,349A only recently.
This explains why these stars are still close to each other.

\subsection{Origin of the equatorial belt of the parsec-scale nebula}
\label{sec:mach}

\begin{figure}
\begin{center}
\includegraphics[width=9cm,angle=0,clip=]{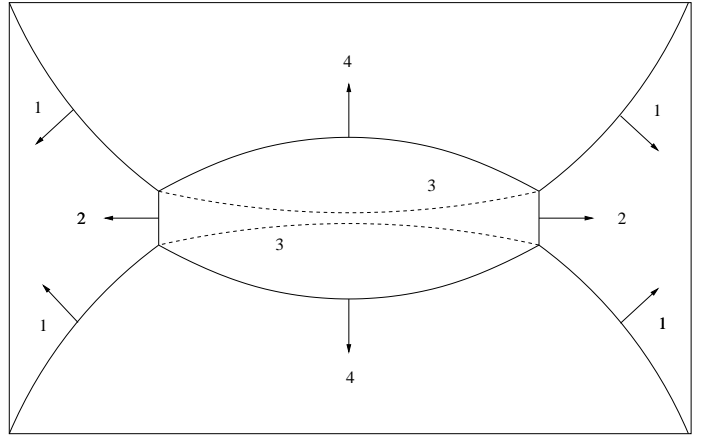}
\end{center}
\caption{Schematic of the proposed origin of the belt around the
 waist of the 24\,$\mu$m bipolar nebula associated with MWC\,349A (not to
 scale): (1) the incident hemispherical shock wave (one of the two lobes of
 the nebula); (2) the ring-like Mach shock wave (the belt); (3) the
 tangential discontinuity; (4) the reflected shock wave. See text for
 details.} \label{fig:mach}
\end{figure}

Now we discuss the possible origin of the belt around the waist of
the mid-infrared nebula associated with MWC\,349A (cf. Gvaramadze
1998). Near the waist each lobe of the bipolar nebula can be
considered as a hemispherical shock wave, which collide with each
other in equatorial plane (see Fig.\,\ref{fig:mach} for schematic
of this interaction). During the collision, the equatorial plane
acts as a wall from which the incident waves are reflected.
Initially, the reflection occurs in a regular fashion: the
incident and reflected shock waves touch the "wall". However, when
the angle between the wall and the tangent to the incident shock
becomes greater than some critical value ($\phi _{\rm cr} \approx
35\degr$ for a medium with an adiabatic index of 5/3), the regular
reflection becomes impossible (e.g. Courant \& Friedrichs 1948).
The incident shock wave is reflected without reaching the
equatorial plane. In addition to the incident and reflected shock
waves, another shock wave, which propagates parallel to the wall,
appears (Fig.\,\ref{fig:mach}). Such a reflection is called a Mach
reflection, and the emerging new shock wave (in our case, in the
shape of a belt; cf. Chernin et al. 1995) is called the Mach shock
wave. The three shock waves intersect along a single (circular)
line, from which the surface of a tangential discontinuity
separates the medium behind the Mach shock wave from the medium
that passes through the incident and reflected ones.

The velocity of the Mach shock wave $v_{\rm M}$ is related to the
velocity of the incident shock wave $v_{\rm i}$ by $v_{\rm M} =
v_{\rm i} /\sin \phi$, where $\phi$ is the angle between the
tangent to the front of the incident shock and the equatorial
plane of the nebula. The Mach wave reaches its maximum velocity at
the beginning of the Mach reflection, i.e., at $\phi = \phi _{\rm
cr}$: $v_{\rm M} ^{\rm max} \approx 1.7v_{\rm i}$. As the angle
$\phi$ increases further, $v_{\rm M}$ decreases, and the Mach wave
degenerates. Using the 24\,$\mu$m image of the nebula, we
estimated that $\phi \approx 50\degr$, so that we expect that the
belt expands with a velocity of $\approx 1.3$ times exceeding the
local lateral expansion velocity of the bipolar lobes.

Our interpretation of the belt around the waist as a Mach shock is
testable. In Sect.\,\ref{sec:Ha} we mentioned that the waist has
an optical counterpart and that its western edge coincides with a
prominent H$\alpha$ filament (Marston \& McCollum 2008). If the
belt is indeed the Mach shock then we expect that high-resolution
spectroscopy of the H$\alpha$ nebula would show that the expansion
velocity of the belt is higher than that of the adjacent parts of
the lobes.

\section{Summary}
\label{sec:sum}

We have reported the discovery of a parsec-scale bipolar nebula
around the curious emission-line star MWC\,349A. The morphology of
the nebula and its orientation are similar to those of the
well-known ($\sim 400$ times smaller) radio nebula associated with
MWC\,349A. The close similarity of the two nebulae suggests that
they are shaped by the same agent -- the disk around MWC\,349A.
Significant photometric and spectroscopic variability of MWC\,349A
along with the presence of the parsec-scale bipolar nebula
strongly argue that this star is an LBV. We have discussed the
physical relationship between MWC\,349A and the nearby B0\,III
star MWC\,349B and suggested that both stars were members of a
hierarchical triple system, which was dynamically ejected from the
core of the Cyg\,OB2 association several Myr ago. The stellar wind
mass loss from the most massive ($\sim 40 \, \msun$) star in the
triple (MWC\,349A) made the system unstable, so that it dissolved
into a binary (MWC\,349A) and a single unbound star (MWC\,349B).
The binary nature of MWC\,349A is supported by the existence of a
9-year periodicity in the red light variations of this star. We
proposed that the gravitational focusing of the slow (red
supergiant) wind from MWC\,349A by the companion star led to the
origin of a circumbinary disk, which currently collimates the fast
(LBV) wind of MWC\,349A and is responsible for the B[e] morphology
of the stellar spectrum.

\begin{acknowledgements}
We are grateful to N.Langer and J.Pflamm-Altenburg for useful
discussions, and to the referee for his comments. This work has
made use of the NASA/IPAC Infrared Science Archive, which is
operated by the Jet Propulsion Laboratory, California Institute of
Technology, under contract with the National Aeronautics and Space
Administration, the SIMBAD database and the VizieR catalogue
access tool, both operated at CDS, Strasbourg, France.
\end{acknowledgements}


\begin{thebibliography}{}

\bibitem{} Altenhoff, W.J., Strittmatter, P.A., \& Wendker, H.J. 1981, A\&A, 93, 48
\bibitem{} Andrillat, Y., Jaschek, M., \& Jaschek, C. 1996, A\&AS, 118, 495
\bibitem{} Baldwin, J.E., Stella Harris, C., \& Ryle, M. 1973, Nature, 241, 39
\bibitem{} Blaauw, A., 1961, Bull. Astron. Inst. Netherl., 15, 265
\bibitem{} Bohannan, B. 1997, in ASP Conf. Ser., Vol. 120, Luminous Blue Variables: Massive Stars in Transition, ed. A.
Nota \& H.J.G.L.M. Lamers (San Francisco: ASP), 120
\bibitem{} Braes, L.L.E., Habing, H.J., \& Schoenmaker, A.A. 1972, Nature, 240, 230
\bibitem{} Brandner, W., Grebel, E.K., Chu, Y.-H., \& Weis, K. 1997, ApJ, 475, L45
\bibitem{} Chernin, A.D., Efremov, Yu.N., \& Voinovich, P.A. 1995, MNRAS, 275, 313
\bibitem{} Ciatti, F., D'Odorico, S., \& Mammano, A. 1974, A\&A, 34, 181
\bibitem{} Clark, J.S., Larionov, V.M., \& Arkharov A. 2005, A\&A, 435, 239
\bibitem{} Cohen, M., Bieging, J.H., Welch, W.J., \& Dreher, J.W. 1985, ApJ, 292, 249
\bibitem{} Comer\'on, F., \& Pasquali, A. 2007, A\&A, 467, L23
\bibitem{} Corradi, R.L.M., \& Schwarz, H.E. 1993, A\&A, 268, 714
\bibitem{} Courant, R., \& Friedrichs, K.O. 1948, Supersonic Glow and Shock Waves (New York: Interscience Publishers)
\bibitem{} Danchi, W.C., Tuthill, P.G., \& Monnier, J.D. 2001, ApJ, 562, 440
\bibitem{} de Zeeuw, P.T., Hoogerwerf, R., de Bruijne, J.H.J., Brown, A.G.A., \& Blaauw, A. 1999, AJ, 117, 354
\bibitem{} de Wit, W.J., Testi, L., Palla, F., \& Zinnecker, H. 2005, A\&A, 437, 247
\bibitem{} Dreher, J.W., \& Welch, W.J. 1983, AJ, 88, 1014
\bibitem{} Egan, M.P., Clark, J.S., Mizuno, D.R., Carey, S.J., Steele, I.A., \& Price, S.D., 2002, ApJ, 572, 288
\bibitem{} Eggleton, P. 2005, Evolutionary processes in binary and multiple stars (Cambridge: Cambridge University Press)
\bibitem{} Eggleton, P., \& Kiseleva, L. 1995, ApJ, 455, 640
\bibitem{} Elvius, A. 1974, A\&A, 34, 371
\bibitem{} Fabian, A.C., \& Hansen, C.J. 1979, MNRAS, 187, 283
\bibitem{} Fazio, G.G. et al. 2004, ApJS, 154, 10
\bibitem{} Frank, A., Balick, B., \& Davidson, K. 1995, ApJ, 441, L77
\bibitem{} Gies, D. R. 1987, ApJS, 64, 545
\bibitem{} Gordon, M.A. 1992, ApJ, 387, 701
\bibitem{} Gordon, M.A. 2003, ApJ, 589, 953
\bibitem{} Gordon, M.A., Holder, B.P., Jisonna, L.J., Jorgenson, R.A., \& Strelnitski, V.S. 2001, ApJ, 559, 402
\bibitem{} Gottlieb, E.W., \& Liller, W. 1978, ApJ, 225, 488
\bibitem{} Gvaramadze, V.V. 1998, Astronomy Letters, 24, 144
\bibitem{} Gvaramadze, V.V., \& Bomans, D.J. 2008a, A\&A, 485, L29
\bibitem{} Gvaramadze, V.V., \& Bomans, D.J. 2008b, A\&A, 490, 1071
\bibitem{} Gvaramadze, V.V., Kniazev, A.Y., \& Fabrika, S. 2010a, MNRAS, 405, 1047
\bibitem{} Gvaramadze, V.V., Kroupa, P., \& Pflamm-Altenburg, J. 2010b, A\&A, 519, A33
\bibitem{} Gvaramadze, V.V., Kniazev, A.Y., Fabrika, S., Sholukhova, O., Berdnikov L.N., Cherepashchuk,
A.M., \& Zharova, A.V. 2010c, MNRAS, 405, 520
\bibitem{} Gvaramadze, V.V. et al. 2009, MNRAS, 400, 524
\bibitem{} Hamann, F., \& Simon, M. 1986, ApJ, 311, 909
\bibitem{} Hamann, F., \& Simon, M. 1988, ApJ, 327, 876
\bibitem{} Hanson, M.M. 2003, ApJ, 597, 957
\bibitem{} Hartmann, L., Jaffe, D., \& Huchra, J.P. 1980, ApJ, 239, 905
\bibitem{} Herzog, A.D., Gehrz, R.D., \& Hackwell, J.A. 1980, ApJ, 236, 189
\bibitem{} Hills, J.G. 1975, AJ, 80, 809
\bibitem{} Hoffer, J.B. 1983, AJ, 88, 1420
\bibitem{} Hofmann, K.-H., Balega, Y., Ikhsanov, N.R., Miroshnichenko, A.S., \& Weigelt, G. 2002, A\&A, 395, 891
\bibitem{} Hoogerwerf, R., de Bruijne, J.H.J., \& Zeeuw, P.T. 2001, A\&A, 365, 49
\bibitem{} Hora, J.L. et al. 2008, New Light on Young Stars: Spitzer's View of Circumstellar
Disks (http://www.ipac.caltech.edu/spitzer2008/proceedings.html)
\bibitem{} Humphreys, R.M. 1991, in Wolf Rayet Stars, ed. K.A. van der Hucht \& B.
Hidayat (Dordrecht: Kluwer), 485
\bibitem{} Humphreys, R.M., \& Davidson, K. 1994, PASP, 106, 1025
\bibitem{} Jim\'{e}nez-Esteban, F.M., Rizzo, J.R., \& Palau, A. 2010, ApJ, 713, 429
\bibitem{} Jorgenson, R.A., Kogan, L.R., \& Strelnitski, V. 2000, AJ, 119, 3060
\bibitem{} Kelly, D.M., Rieke, G.H., \& Campbell, B. 1994, ApJ, 425, 231
\bibitem{} Kiminki, D.C. et al. 2007, ApJ, 664, 1102
\bibitem{} Kiseleva, L.G., Eggleton, P.P., \& Orlov, V.V. 1994, MNRAS, 270, 936
\bibitem{} Kn\"{o}dlseder, J. 2000, A\&A, 360, 539
\bibitem{} Kraemer, K.E. et al. 2010, AJ, 139, 2319
\bibitem{} Kroupa, P. 1998, MNRAS, 298, 231
\bibitem{} Kroupa, P., \& Boily, C.M. 2002, MNRAS, 336, 1188
\bibitem{} Lada, C.J., \& Lada, E.A. 2003, ARA\&A, 41, 57
\bibitem{} Lamers, H.J.G.L.M., Zickgraf, F.-J., de Winter, D., Houziaux, L., \& Zorec, J. 1998, A\&A, 340, 117
\bibitem{} Langer N., 1997, in ASP Conf. Ser., Vol. 120, Luminous Blue Variables: Massive Stars in Transition, ed.
A. Nota A. \& H. Lamers (San Francisco: ASP), 83
\bibitem{} Langer, N. 1998, A\&A, 329, 551
\bibitem{} Langer, N., Garc\'{i}a-Segura, G., \& Mac Low, M.-M. 1999, ApJ, 520, L49
\bibitem{} Leinert, C. 1986, A\&A, 155, L6
\bibitem{} Livio, M., \& Pringle, J.E. 1998, MNRAS, 295, L59
\bibitem{} Livio, M., Salzman, J., \& Shaviv, G. 1979, MNRAS, 188, 1
\bibitem{} Maeder, A.,  \& Desjacques, V. 2001, A\&A, 372, L9
\bibitem{} Marston, A.P., \& McCollum, B. 2008, A\&A, 477, 193
\bibitem{} Mart\'in-Pintado, J., Bachiller, R., Thum, C., \& Walmsley, M. 1989, A\&A, 215, L13
\bibitem{} Mastrodemos, N., \& Morris, M. 1999, ApJ, 523, 357
\bibitem{} Matthews, L.D., Greenhill, L.J., Goddi, C., Chandler, C.J., Humphreys, E.M.L., \& Kunz, M.W. 2010, ApJ, 708, 80
\bibitem{} Merrill, P.W., Humason, M.L., \& Burwell, C.G. 1932, ApJ, 76, 156
\bibitem{} Meyer, J.M., Nordsieck, K.H., \& Hoffman, J.L. 2002, AJ, 123, 1639
\bibitem{} Mizuno, D.R. et al. 2010, AJ, 139, 1542
\bibitem{} Morse, J.A., Davidson, K., Bally, J., Ebbets, D., Balick, B., \& Frank, A. 1998, AJ, 116, 2443
\bibitem{} Morris, M. 1981, ApJ, 249, 572
\bibitem{} Negueruela, I., Marco, A., Herrero, A., \& Clark, J.S. 2008, A\&A, 487, 575
\bibitem{} Nota, A., Livio, M., Clampin, M., \& Schulte-Ladbeck, R. 1995, ApJ, 448, 788
\bibitem{} Olnon, F.M. 1975, A\&A, 39, 217
\bibitem{} Oudmaijer, R.D., Davies, B., de Wit, W.-J., \& Patel, M. 2009, in ASP Conf. Ser., Vol. 412, Biggest,
Baddest, Coolest Stars, ed. D.G. Luttermoser, B.J. Smith \& R.E.
Stencel (San Francisco: ASP), 17
\bibitem{} Owocki, S.P., \& Gayley, K.G. 1997, in ASP Conf. Ser.,
Vol. 120, Luminous Blue Variables: Massive Stars in Transition,
ed. A. Nota A. \& H. Lamers, (San Francisco: ASP), 121
\bibitem{} Parker, R.J., \& Goodwin, S.P. 2007, MNRAS, 380, 1271
\bibitem{} Pflamm-Altenburg, J., \& Kroupa, P. 2010, MNRAS, 404, 1564
\bibitem{} Piddington, J.H., \& Minnett, H.C. 1952, AuSRA, 5, 17
\bibitem{} Plambeck, R.L., Wright, M.C.H., Friedel, D.N. et al. 2009, ApJ, 704, L25
\bibitem{} Planesas, P., Mart\'in-Pintado, J., \& Serabyn, E. 1992, ApJ, 386, L23
\bibitem{} Ponomarev, V.O., Smith, H.A., \& Strelnitski, V.S. 1994, ApJ, 424, 976
\bibitem{} Reid, M.J., Menten, K.M., Zheng, X.W., Brunthaler, A., \& Xu, Y. 2009, ApJ, 705, 1548
\bibitem{} Reipurth, B., \& Schneider, N., 2008, in ASP Monograph
Publications, Vol. 4, Handbook of Star Forming Regions, Vol. I:
The Northern Sky, ed. B.Reipurth (San Francisco: ASP), 36
\bibitem{} Rieke, G.H. et al. 2004, ApJS, 154, 25
\bibitem{} Rodr\'{i}guez, L.F., G\'{o}mez, Y., \& Tafoya, D. 2007, ApJ, 663, 1083
\bibitem{} Russeil D., 2003, A\&A, 397, 133
\bibitem{} Rygl, K. et al. 2012, A\&A, 539, A79
\bibitem{} Sahai, R. et al. 1999, AJ, 118, 468
\bibitem{} Schilbach, E., \& R\"{o}ser, S. 2008, A\&A, 489, 105
\bibitem{} Schneider, N. et al. 2006, A\&A, 458, 855
\bibitem{} Sch\"{o}nrich, R., Binney, J., \& Dehnen, W. 2010, MNRAS, 403, 1829
\bibitem{} Selier, R., Heydari-Malayeri, M., \& Gouliermis, D.A. 2011, A\&A, 529, A40
\bibitem{} Smith, N. 2007, AJ, 133, 1034
\bibitem{} Strelnitski, V., Haas, M.R., Smith, H.A., Erickson, E.F., Colgan, S.W.J., \& Hollenbach,
D.J. 1996, Science, 272, 1459
\bibitem{} Swings, P., \& Struve, O. 1942, ApJ, 95, 152
\bibitem{} Tafoya, D., G\'{o}mez, Y., \& Rodr\'{i}guez, L.F. 2004, ApJ, 610, 827
\bibitem{} Tanaka, M., Yamashita, T., Sato, S., Nishida, M., Ukita, N., \& Okuda,
H. 1985, PASP, 97, 1115
\bibitem{} Thompson, R.I., Strittmatter, P.A., Erickson, E.F., Witteborn, F.C., \& Strecker, D.W. 1977, ApJ, 218, 170
\bibitem{} Thum, C., Mart\'in-Pintado, J., \& Bachiller, R. 1992, A\&A, 256, 507
\bibitem{} Trams, N.R., Voors, R.H.M., \& Waters, L.B.F.M. 1998, Ap\&SS, 255, 195
\bibitem{} Voors, R.H.M., Geballe, T.R., Waters, L.B.F.M., Najarro, F., \& Lamers, H.J.G.L.M. 2000, A\&A, 362, 236
\bibitem{} Wachter, S., Mauerhan, J.C., van Dyk, S.D., Hoard, D.W., Kafka, S., \& Morris, P.W. 2010, AJ, 139, 2330
\bibitem{} Wachter, S., Mauerhan, J., van Dyk, S., Hoard, D.W., \& Morris, P., 2011, Bull.
Soc. R. Sci. Li\`{e}ge, 80, 291
\bibitem{} Walborn, N.R., \& Fitzpatrick, E.L. 2000, PASP, 112, 50
\bibitem{} Weidner, C., Gvaramadze, V.V., Kroupa, P., Pflamm-Altenburg, J. 2011, in Stellar Clusters and Associations (A RIA
workshop on GAIA, 23-27 May 2011, Granada, Spain), ed. E.J. Alfaro
Navarro, A.T. Gallego Calvente, M.R. Zapatero Osorio, 264
\bibitem{} Weis, K. 2001, RvMA, 14, 261
\bibitem{} Weis, K. 2011, Bull. Soc. R. Sci. Li\`{e}ge, 80, 440
\bibitem{} Weintroub, J., Moran, J.M., Wilner, D.J., Young, K., Rao, R., \& Shinnaga, H. 2008, ApJ, 677, 1140
\bibitem{} White, R.L., \& Becker, R.H. 1985, ApJ, 297, 677
\bibitem{} Yudin, R.V. 1996, A\&A, 312, 234

\end{thebibliography}
\end{document}